\documentclass[rnote]{aa}
\usepackage{txfonts}
\usepackage{graphicx}
\begin{document}

\title{Deep imaging survey of the environment of $\alpha$\,Centauri}
\subtitle{II. CCD imaging with the NTT-SUSI2 camera}
\titlerunning{CCD survey of the environment of $\alpha$ Centauri}
\authorrunning{P. Kervella \& F. Th\'evenin}
\author{
P.~Kervella\inst{1}
\and
F. Th\'evenin\inst{2}
}
\offprints{P. Kervella}
\mail{Pierre.Kervella@obspm.fr}
\institute{LESIA, UMR 8109, Observatoire de Paris-Meudon, 5, place Jules Janssen, 
F-92195 Meudon Cedex, France
\and Laboratoire Cassiop\'ee, Observatoire de la C\^ote d'Azur, BP 4229, F-06304 Nice Cedex 4, France
}
\date{Received ; Accepted}
\abstract
{Thanks to its proximity, $\alpha$\,Centauri is an outstanding target for an imaging
search for extrasolar planets.
}
{We searched for faint comoving companions to $\alpha$\,Cen located at angular distances
of a few tens of arcseconds, up to 2-3\,arcmin.}
{We obtained CCD images from the NTT-SUSI2 instrument in the Bessel
$V$, $R$, $I$, and $Z$ bands, and archive data
from 2MASS.}
{We present a catalogue of the detected objects inside a 5.5\,arcmin box around this star.
A total of 4313 sources down to $m_V \approx 24$ and $m_I \approx 22$
were detected in the SUSI2 images.
The infrared photometry of part of these sources has been extracted from 
the 2MASS images.}
{No comoving companion to $\alpha$\,Centauri were detected between 100 and 300\,AU,
down to a maximum mass of $\approx$15 times Jupiter.
We also mostly exclude the presence of a companion more massive than
30\,M$_J$ between 50 and 100\,AU.}


\maketitle
%
\section{Introduction}

Our closest stellar neighbour, the $\alpha$\,Cen visual triple star ($d = 1.34\,$pc),
is an extremely attractive target for extrasolar planet detection. The main
components $\alpha$\,Cen\,A (\object{HD 128620}) and B (\object{HD 128621})
are G2V and K1V solar-like stars, while the third member is the red dwarf Proxima (M5.5V).
In all imaging planet searches, the main difficulty is to retrieve the
planetary signal in the bright diffuse halo from the star.
As discussed in Paper~I (Kervella et al.~\cite{kervella06}), there are indications
that the gravitational mass of $\alpha$\,Cen~B could be higher than its modeled mass.
For these reasons, we engaged in a search for companions to $\alpha$\,Cen
at two different angular scales: a few arcseconds, using adaptive optics (Paper~I),
and a few tens of arcseconds, using CCD imaging (present work).
To our knowledge, this is the first wide-field search
for companions to $\alpha$\,Cen.

\section{Observations \label{observations}}

Very close to the two stars, within a radius of about 20", adaptive optics imaging allows
to reach the best sensitivity. At distances of more than 20", the diffused light is less of a problem, and
atmosphere limited imaging is most cost-effective solution.
We thus observed $\alpha$\,Cen using ESO's 3.6\,m NTT,
equipped with the SUperb Seeing Imager 2 (SUSI2).
This instrument is based on a
mosaic of two 2k$\times$4k thinned EEV CCDs (15\,$\mu$m pixels).
In order to avoid an heavy saturation of the detector, we positioned the $\alpha$\,Cen
pair aligned within the 8\,arcsec gap between the two detectors. The Nasmyth
adapter of the telescope was rotated to align the two stars with the direction of the gap.
The pixel scale of SUSI2 is 0.0805\,"/pixel, therefore providing an excellent
sampling of the point spread function.
We obtained a series of exposures through four filters, Bessel $V$, $R$, $I$, and
$Z$\footnote{The transmission curves of the SUSI2 filters are
available from http://www.ls.eso.org/lasilla/sciops/ntt/susi/docs/SUSIfilters.html},
in order to be able to confirm potential companions based on their colors.
The individual exposures times were short, typically 30 to 60\,s, to limit the
saturation of the detector to a small area around $\alpha$\,Cen. The journal of the
SUSI2 observations is presented in Table~\ref{susi2_log}. The total shutter open
time obtained on $\alpha$\,Cen with SUSI2 in all bands reaches almost 4\,hours.

\begin{table}
\caption{Log of the SUSI2 observations of $\alpha$\,Cen. The indicated UT time
corresponds to the middle of the exposures, $\theta$ is the FWHM of the point spread function
of the composite images, and AM is the airmass. The observations of 26 Feb 2004 were excluded
due to poor seeing.} 
\label{susi2_log}
\begin{small}
\begin{tabular}{lccccc}
\hline
Date & UT & Filter & Total exp. (s) & $\theta$\,(") & AM \\
\hline
2004-02-25 & 6:40 & $R$ & 1185 & 0.74 & 1.30 \\
2004-02-26 & 7:48 & $V$ & 555 & 2.67 & 1.20 \\
2004-02-29 & 6:31 & $I$ & 2010 & 0.81 & 1.29 \\
2004-02-29 & 7:54 & $V$ & 1305 & 0.84 & 1.19\\
2004-04-02 & 2:55 & $Z$ & 2360 & 0.89 & 1.51 \\
2004-04-02 & 4:21 & $V$ & 1305 & 1.01 & 1.29 \\
2004-04-02 & 5:49 & $I$ & 2010 & 0.97 & 1.19 \\
\hline
2006-04-11 & 8:29 & $I$ & 1649 & 1.01 & 1.30 \\
2006-08-17 & 1:22 & $I$ & 1649 & 0.65 & 1.49 \\
\hline
\end{tabular}
\end{small}

\end{table}

In order to obtain infrared photometry for part of the sources detected in the SUSI2 images,
we retrieved from the IPAC archive\footnote{http://irsa.ipac.caltech.edu/} the FITS images
from the 2MASS survey (Skrutskie et al.~Ê\cite{skrutskie06})
covering the field of view of our SUSI2 images.The selected images
were all obtained in early 2000.

\section{Data reduction \label{processing}}

\subsection{Raw data processing and astrometry}

\begin{figure*}
\centering
\includegraphics[width=12.5cm, angle=0]{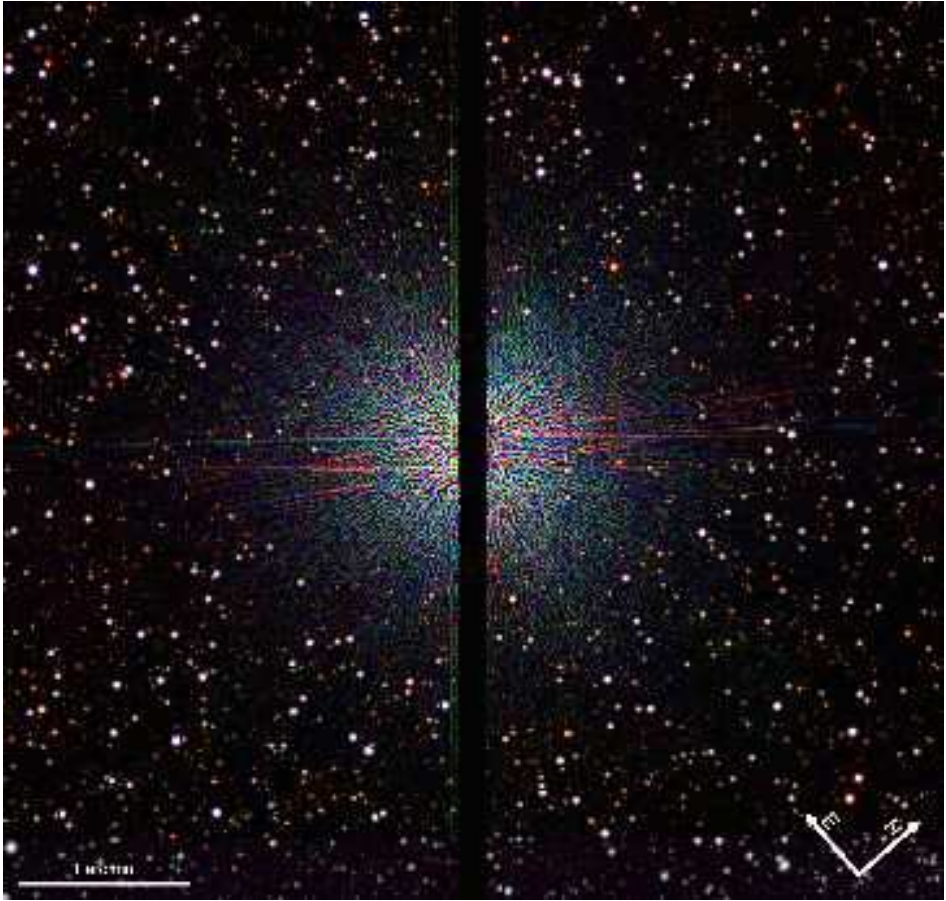}
\caption{Tricolor filtered image from the SUSI2 images of $\alpha$\,Cen (February-April 2004).}
\label{tricolor_susi2}
\end{figure*}

We applied the standard CCD image processing steps to our SUSI2 images,
using IRAF: bias subtraction, flat-fielding, bad pixel masking.
The 2MASS images are available fully calibrated from the IPAC archive.
The resulting SUSI2 band image is presented in Fig.~\ref{tricolor_susi2},
where the $VRI$ images have been mapped to the blue, green and red colors.
The lower $\approx$10\% area of this image is shown in grey levels as only
two colors are available.
The astrometric calibration of our SUSI2 images has been obtained by matching the
coordinates of the brightest sources with their counterparts in the 2MASS
images. The astrometric accuracy of the 2MASS images is about 0.1"
(Skrutskie et al.~\cite{skrutskie06}). Due to the presence of stronger diffused light in the visible, 
we estimate our astrometric uncertainty at 0.2" over the
SUSI2 field, therefore this astrometric calibration method is sufficient for our purpose.

\subsection{Sources extraction}

The SUSI2 sources were extracted using the SExtractor tool (Bertin \& Arnouts~\cite{bertin96}),
completed by a manual selection of the sources close to $\alpha$\,Cen.
We applied the same hybrid selection procedure on the 2MASS images.
We then obtained the photometry of the sources using IRAF
and 12\,pixel apertures for SUSI2 (1" radius). This relatively small
aperture is justified by the variable background from the diffused light of $\alpha$\,Cen and
the relative crowding of the field. For the 2MASS images, we used 3" apertures.
The SUSI2 airmass coefficients were taken as the average values for La Silla
(ESO User's Manual~1993, see also Burki et al.~\cite{burki95}):
$k_V=0.11$, $k_R=0.07$, $k_I=0.02$, and $k_Z=0.01$\,mag/airmass.
In all cases, the airmass corrections are negligible, with values below 0.04\,mag.
The aperture corrections for SUSI2 were determined
on a bright, isolated star located relatively far from $\alpha$\,Cen to ensure a flat sky
background.
They were found to range between 0.20 and 0.42\,mag depending on the color and the
observation epoch. The photometric zero points were taken from ESO's routine
instrument monitoring program for SUSI2, and from the image headers for 2MASS.
Considering the difficulty of estimating accurately the sky background level, we chose to add a
systematic $\pm 0.20$\,mag uncertainty to the SUSI2 magnitudes.
For the 2MASS images, we applied the aperture correction recommended by IPAC
for a 3" aperture, i.e. $0.05 \pm 0.02$\,mag in all three $JHK$ bands.

The catalogue of the objects detected in the SUSI2 and 2MASS images is 
available in electronic format\footnote{http://vizier.u-strasbg.fr/viz-bin/VizieR}.
A total of 4313 sources were identified on the SUSI2 images, with 391 recovered in the 2MASS
images (9.1\%). Among the detected background sources,
several objects present extreme color indices. We checked that their presence is expected based
on Galactic models (Robin et al.~\cite{robin03}).

\section{Sensitivity and companion limits \label{sensitivity}}

To define our practical limiting magnitudes,
we chose to consider the median magnitude of the detected sources.
This definition has the advantage to give an empirical, statistically meaningful
definition of the sensitivity, that can be expressed as a function of the distance to the
two bright stars by computing the median within angular distance bins.
We computed the median magnitude of the detected objects in the
SUSI2 $VRIZ$ bands for four angular distance bins:
$23-50$", $50-100$", $100-150$", $150-200$" and $200-236$".
In all bands, the limiting magnitudes (as defined above)
at large angular distances are in the 20-22 range,
with the faintest objects at magnitudes of 23-25.
The limiting magnitudes decrease to 16-18 for an augular separation of 35".
For the 2MASS images, the limiting magnitudes as defined above
are13-14 for the $JHK$ bands at large angular separations (the faintest sources
are at 15-16).

\begin{figure}
\centering
\includegraphics[bb=0 0 360 180, width=8.5cm]{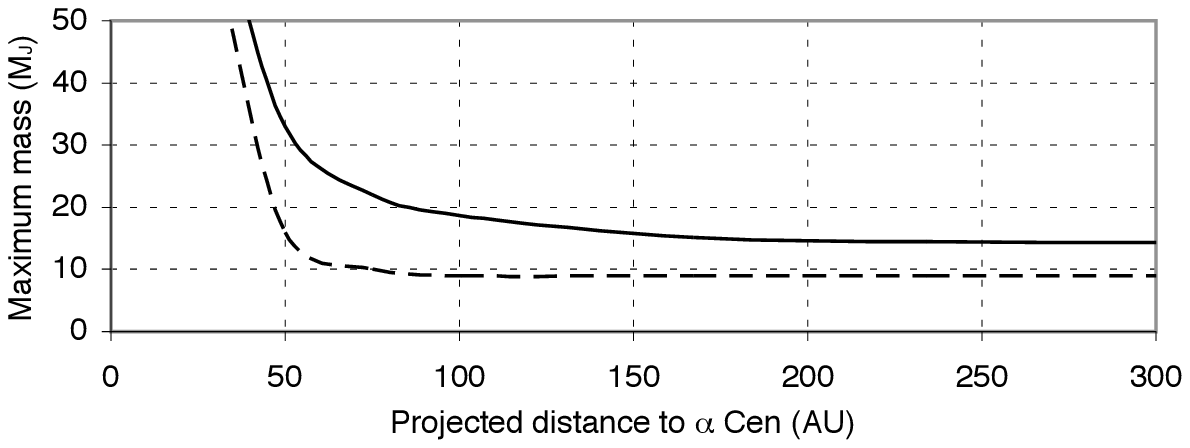}
\caption{Maximum mass of the possible companions orbiting $\alpha$\,Cen.
The solid curve corresponds to the model masses by Baraffe et al.~(\cite{baraffe03})
for our SUSI2 limiting magnitudes in the $I$ band.
The dashed curve corresponds to the faintest detected objects.}
\label{companions}
\end{figure}

We carefully searched our SUSI2 images for moving sources using
blinking and image subtraction techniques, without success.
Using planetary models from Baraffe et al.~(\cite{baraffe03}) for the age
of $\alpha$\,Cen (5\,Gyr, from Th\'evenin et al.~\cite{thevenin02}, see also
Eggenberger et al.~\cite{eggenberger04}), we can estimate
the maximum mass of the possible companions.
The most constraining limits are provided by our $I$ band images.
The $Z$ filter did not provide a significant improvement in sensitivity, due to the
cutoff of the CCD quantum efficiency. The $I$ band provides in addition a better defined
average wavelength that makes the comparison with model magnitudes more accurate.
Fig.~\ref{companions} gives the sensitivity of our search in terms of
companion mass. We can set a maximum mass of $\approx 15$\,M$_J$
for projected separations larger than 100\,AU. This limit can be extended down to 10\,M$_J$
if we consider our faintest detected objects. Between 50 and 100\,AU, the presence of
a companion with a mass larger than about 30\,M$_J$ is highly unlikely.

\section{Conclusion}

From deep CCD images of $\alpha$\,Cen, we did not identify any
comoving companion. Within the explored area, this negative result sets an upper
mass limit of 15-30\,M$_J$ to the possible companions orbiting $\alpha$\,Cen~B
or the pair, for separations of 50-300\,AU. When combined with existing
radial velocity searches (e.g. Endl et al.~Ê\cite{endl01}) and our adaptive optics results
(see Fig.~7 in Paper~I), this mostly excludes the presence of a 20-30\,M$_J$ companion
within 300\,AU.
This non-detection should have consequences on the modeling
of the stellar interior of $\alpha$\,Cen in particular, but also of
other stars. In order to settle the debate on the masses of  $\alpha$\,Cen
(and in particular of B), one should now check one by one
the hypotheses made for the calibration of this system.
This also implies that the new abundances of Asplund et al. (\cite{asplund04}) should be
considered as soon as possible.
As an illustration of the importance of this parameter, it has been debated
recently a possible underestimate of the Ne solar abundance that could counter
balance the effect of the decrease of the solar opacity due to the decrease of
the oxygen abundance from 3D models (Young~\cite{young05}). Liefke
\& Schmitt~(\cite{liefke06}) have published a new value of the
chemical abundance ratio of Ne/O based on coronal studies.
These recent results will be advantageously introduced in the determination of the
metallicity of the $\alpha$\,Cen binary to better model the evolutionnary track of
both stars.

\begin{acknowledgements}
Based on observations made with ESO Telescopes at La Silla
under programs 272.C-5010 and 077.C-0587.
\end{acknowledgements}

{}
\end{document}